\documentclass[twocolumn,showpacs,aps,ajp]{revtex4}

\usepackage{amsmath}
\usepackage{amssymb}
\usepackage{graphicx}
\usepackage{graphics}

\begin{document}

\title{Legendre transforms of the fundamental thermodynamic relation and statistical ensembles}
\author{S. Stepanow}
\affiliation{Martin-Luther-Universit\"{a}t Halle-Wittenberg, Institut f\"{u}r Physik,
D-06099 Halle, Germany }
\date{\today}

\begin{abstract}
We show how the Legendre transforms of the fundamental thermodynamic relation can be used  to introduce different
statistical ensembles.
\end{abstract}

\pacs{05.}
\maketitle


\section{ Introduction}

\label{intro}

The thermodynamic potentials  can be introduced using the Legendre transform of the fundamental thermodynamic relation.
Because some of the thermodynamic potentials appear as normalization factors in different statistical
ensembles, e.g. the free energy in the canonical ensemble, it arises the question how the Legendre transforms can be used
to introduce corresponding statistical ensembles. We will give in the present article an answer to this question.
The procedure of introduction of statistical ensembles which will be presented in this article is an alternative to the standard treatment of statistical ensembles \cite{landau-lifschitz5}. This procedure was used in \cite{stepanow09} to establish the partition function in an external electric field at constant potential. It can be also used in teaching of statistical physics.

\section{Legendre transforms and statistical ensembles}\label{formal}

The Legendre transforms to change the independent variables in the fundamental thermodynamic relation of thermodynamics \cite{landau-lifschitz5}
\begin{equation}\label{st2}
    dU=TdS-pdV+\mu dN
\end{equation}
enables one to introduce the thermodynamic potentials.
The free energy and the Gibbs potential  are derived as follows
\begin{equation}\label{st3}
   d(U-TS)=dF = -SdT-pdV+\mu dN,
\end{equation}
\begin{equation}\label{st4}
    d(U-TS+pV)=dG = -SdT+Vdp+\mu dN,
\end{equation}
where $G=\mu N$ applies for one component system.

We now will show how to introduce the grand canonical ensemble by starting with the relation between the free energy and the canonical partition function
\begin{equation}\label{st6}
    e^{\beta F(T,V,N)} Z(T,V,N)=1,
\end{equation}
where $\beta=1/k_B T$.
We replace $F$ in (\ref{st6}) using the relation $G=F+pV=F-\Omega$
by $F=\Omega+\mu N$ with $\Omega=-pV$ and arrive at the relation in the canonical ensemble
\begin{equation}\label{st7}
    e^{\beta \Omega+\beta\mu N}Z(T,V,N)=1,
\end{equation}
which is fulfilled for fixed number of particles $N$.
In the ensemble with not fixed number of particles the expression $e^{\beta\Omega}e^{\beta\mu N}Z(T,V,N)$ is likely to be interpreted as the statistical weight of the subsystem with a particular number of particles $N$.
Therefore, in the statistical ensemble with arbitrary $N$ we demand instead of Eq.~(\ref{st7}) the fulfillment of the condition
\begin{equation}\label{st8}
    \sum_{N=0}^{\infty}e^{\beta \Omega+\beta\mu N}Z(T,V,N)=1,
\end{equation}
which results in the following expression for the grand canonical partition function
\begin{equation}\label{st9}
    \Xi(T,V,\mu)=e^{-\beta\Omega}=\sum_{N=0}^{\infty}e^{\beta\mu N}Z(T,V,N).
\end{equation}
The transition from Eq.~(\ref{st7}), which is the consequence of the Legendre transform, to Eq.~(\ref{st8}) is an alternative to the standard derivation of the grand canonical partition function \cite{landau-lifschitz5}.
The average number of particles in the grand canonical ensembles can be derived from Eq.~(\ref{st9}) as follows
\begin{equation}\label{st10}
    <N>=-\left(\frac{\partial\Omega}{\partial \mu}\right)_{T,V}.
\end{equation}
Eq.~(\ref{st10}) together with the relation $\Omega =-pV$ build the basis for calculation of thermodynamic quantities in the grand canonical ensemble.

As the next example, we will show how to introduce the canonical ensemble by starting with the relation in the micro canonical
ensemble
\begin{equation}\label{mcc1}
    e^{-S/k_B}\Delta\Gamma(E)=1,
\end{equation}
where $\Delta\Gamma(E)$ is the number of states of the system at the energy $E$.
Using the thermodynamic relation $ S=(E-F)/T$ we arrive at
\begin{equation}\label{mcc3}
    e^{\beta F}e^{-\beta E}\Delta\Gamma(E)=1.
\end{equation}
Further, we proceed as above and demand that in the ensemble with not fixed $E$ the relation (\ref{mcc3}) is fulfilled on
the average i.e. as integral of the expression on the left-hand side over the energy (sum for discrete states). As a result we arrive at the following expression
for the canonical partition function
\begin{equation}\label{mcc4}
    Z(T,V,N)=e^{-\beta F}=\int_0^{\infty}dE e^{-\beta E}\Delta\Gamma(E).
\end{equation}
The above procedure works for all possible ensembles.

In the following we will consider the more non-trivial example of the realization of the above idea
to derive the partition function in the external electric field at constant potential \cite{stepanow09}.
The free energies  at constant dielectric displacement $F(T,V,\mathbf{D})$ and at constant electric field
$\tilde{F}(T,V,\mathbf{E})$ are related by a Legendre transform as $\tilde{F}=F-\mathbf{ED}V/4\pi $ with the
differentials given by \cite{landau-lifshitz8}
\begin{eqnarray} \label{td2}
dF&=&-SdT-pdV+\frac{V}{4\pi }\mathbf{E}d\mathbf{D},  \notag \\
d\tilde{F}&=&-SdT-pdV-\frac{V}{4\pi }\mathbf{D}d\mathbf{E},
\end{eqnarray}
where for simplicity homogeneous fields are assumed in the above expressions.

We will now consider the question of the statistical mechanical calculation
of the thermodynamic quantities in the ensemble at constant potential.
Proceeding as above we start with the relation
\begin{equation}\label{td3}
    e^{\beta F}Z_{N}(\mathbf{E}_{0})=1,
\end{equation}
where $Z_{N}(\mathbf{E}_{0})$ is the partition function in the external electric field $\mathbf{E}_{0}$
given by
\begin{equation}\label{vw2}
Z_{N}(E_0)=\int d\Gamma e^{-\beta H_{0}-\beta H_{pol}},
\end{equation}%
where $d\Gamma$ denotes here integrations over the phase space, and $H_{0}$ the Hamiltonian at $\textbf{E}_0=0$,
and the total interaction energy of the induced dipoles $H_{pol}$ is given by \cite{isihara62}
\begin{eqnarray}\label{e12}
H_{pol} &=&-\frac{1}{2}\sum\limits_{i,j}E_{0}(r_{i})\left[ 1+\alpha _{i}T%
\right] _{ij}^{-1}\alpha _{j}E_{0}(r_{j})  \notag \\
&=&-\frac{1}{2}\sum\limits_{i}p_{i}^{\alpha }E_{0}^{\alpha }(r_{i})
\end{eqnarray}
with
\begin{equation*}
T^{\alpha \beta }(\mathbf{r}_{j}-\mathbf{r}_{i})=-\nabla _{j}^{\alpha
}\nabla _{j}^{\beta }\frac{1}{r_{ji}}=\frac{\delta ^{\alpha \beta }}{%
r_{ji}^{3}}-3\frac{r_{ji}^{\alpha }r_{ji}^{\beta }}{r_{ji}^{5}}
\end{equation*}
being the tensor of dipole-dipole interactions. The molecular
polarization tensor $\alpha_i $ is defined by $p_{i}^{\alpha }=\alpha
_{i}^{\alpha \beta }E_{0}^{\beta }(\mathbf{r}_{i})$ with $a_i^{\alpha \beta }=\alpha \delta ^{\alpha \beta }$.
Further we replace $F$ in (\ref{td3}) by
\begin{equation}\label{td4}
    F=\tilde{F}+\int d^3 r\mathbf{ED}/4\pi = \tilde{F}+\int d^3 r\mathbf{EE_0}/4\pi ,
\end{equation}
add the energy of the external electric field $\int d^{3}r\mathbf{E}_{0}^{2}/8\pi $, integrates over $\mathbf{E}_0(\mathbf{r})$, and arrive at the following expression for the configurational integral
\begin{eqnarray} \label{td5}
\tilde{Z}_{N}(\mathbf{E})&=&\exp(-\beta\tilde{F}) =\int D\mathbf{E}_{0}(\mathbf{r})\exp \left(
\beta \int d^{3}r\frac{\mathbf{E}\mathbf{E}_{0}}{4\pi }\right)  \notag \\
&\times &\exp \left( -\beta \int d^{3}r\frac{\mathbf{E}_{0}^{2}}{8\pi }%
\right) Z_{N}(\mathbf{E}_{0}).
\end{eqnarray}%
The external electric field $\mathbf{E}_{0}$ is identified in (\ref{td4})
with the dielectric displacement $\mathbf{D}$, because both obey the same
Maxwell equation $\mathrm{div}\mathbf{E}_{0}=4\pi \rho _{ext}$.
The integration over the field strength in (\ref{td5}) occurs at every $\mathbf{r}$, i.e. (\ref{td5}) is a functional integral.

As an application of (\ref{td5}) we will derive the Clausius-Mossotti relation \cite{stepanow09}.
The functional integral in (\ref{td5}) is Gaussian, and consequently
the integration over $\mathbf{E}_{0}(\mathbf{r})$ can be performed exactly using the quadratic complement. We obtain
\begin{eqnarray} \label{td5c}
&&\tilde{Z}_{N}(\mathbf{E})= \int d\Gamma_q \exp \left( -\beta H_{0}\right)
\exp \left[ \frac{1}{2}\ln (\frac{8\pi ^{2}}{\beta }\det A^{-1})\right]
\notag \\
&\times &\exp \left[ \frac{\beta }{8\pi }\int d^{3}r\int d^{3}r^{\prime
}E(r)A^{-1}(r,r^{\prime })E(r^{\prime })\right] ,
\end{eqnarray}%
where $d\Gamma_q $ denotes here integrations over positions $r_{1},\cdots
,r_{N}$ of the particles, and the matrix $A$ is defined by
\begin{equation} \label{td5d}
A(r,r^{\prime })=\delta (r-r^{\prime })-4\pi n(r)(I+\alpha Tn)_{r,r^{\prime
}}^{-1}\alpha ,
\end{equation}
where $I\rightarrow \delta (r-r^{\prime })$ is the identity matrix, and $n(r)
$ is the microscopic density $n(r)=\sum\limits_{i}\delta (r-r_{i})$. The
Cartesian indices of $A$ are suppressed.
The preaveraging of $A(r,r^{\prime })$ in the exact expression (\ref{td5d}) according to
\begin{eqnarray}\label{ef1}
    A(r,r^{\prime })&\rightarrow& \langle A(r,r^{\prime })\rangle \nonumber \\
    &=& \delta (r-r^{\prime })\left( 1-4\pi n\alpha \left( 1+8\pi \alpha n/3\right)
^{-1}\right), \nonumber \\
\end{eqnarray}
where $\alpha$ is the polarizability, results in the following expression \cite{stepanow09}
\begin{eqnarray}\label{ef2}
    \exp(-\beta\tilde{F})
     = \exp\left(-\beta F_0 +\beta\int\varepsilon \mathbf{E}^2(\mathbf{r})d^3r/8\pi\right)
\end{eqnarray}
with the dielectric constant
\begin{equation}\label{ef3}
    \frac{1}{\varepsilon} =1-\frac{4\pi n\alpha}{1+8\pi n\alpha /3} ,
\end{equation}
which is equivalent to the Clausius-Mossotti relation.

%

\end{document}